\documentclass[allclo]{FBSart}
\usepackage{amsfonts}
\usepackage{amssymb}
\usepackage{epsfig}
\usepackage{bm}
\def\dis{\displaystyle}
\def\itmb{\begin{itemize}}
\def\itme{\end{itemize}}
\def\enmb{\begin{enumerate}}
\def\enme{\end{enumerate}}
\def\eqnb{\begin{equation}}
\def\eqne{\end{equation}}
\def\PTP{Prog. Theor. Phys.(Kyoto)}

\def\NPB{{Nucl. Phys.} {\bf B}}
\def\PLB{{Phys. Lett.} B}
\def\PRL{Phys. Rev. Lett.}
\def\PRD{{Phys. Rev.} D}

\title{Infrared features of unquenched Lattice Landau Gauge QCD}

\author{Sadataka Furui \instnr{1}\thanks{\textit{E-mail address:} furui@umb.teikyo-u.ac.jp} and 
Hideo Nakajima \instnr{2}\thanks{\textit{E-mail address:} nakajima@is.utsunomiya-u.ac.jp}}
\institute{\instnr{1} School of Science and Engineering, Teikyo University, 320-8551, Utsunomiya, Japan,

\instnr{2} Department of Information Science, Utsunomiya University, 321-8585, Utsunomiya, Japan. }

\runningauthor{S.\,Furui and H.\,Nakajima}
\runningtitle{Infrared features of unquenched Lattice Landau Gauge QCD}
\begin{document}

\maketitle

\begin{abstract}
The running coupling and the Kugo-Ojima parameter of unquenched lattice Landau gauge are simulated and compared with the continuum theory. Although the running coupling measured by the ghost and gluon dressing function is infrared suppressed, the running coupling has the maximum of $\alpha_0\sim 2-2.5$ at around $q=0.5$GeV irrespective of the fermion actions (Wilson fermions and Kogut-Susskind(KS) fermions). The Kugo-Ojima parameter $c$ which saturated to about 0.8 in quenched simulations becomes consistent with 1 in the MILC configurations produced with the use of the Asqtad action, after averaging the dependence on polarization directions caused by the asymmetry of the lattice. Presence of $1+c_1/q^2$ correction factor in the running coupling depends on the lattice size and the sea quark mass. In the large lattice size and small sea quark mass, $c_1$ is confirmed of the order of a few GeV. The MILC configuration of $a=0.09$fm suggests also the presence of dimension-4 condensates with a sign opposite to the dimension-2 condensates. 
The gluon propagator, the ghost propagator and the running coupling are compared with recent pQCD results including an anomalous dimension of fields up to the four-loop level.
\end{abstract}

\section{Introduction}
In 1978 Gribov showed that the Landau gauge fixing of the Yang-Mills theory does not define the unique gauge but there appear gauge equivalent copies and that the sufficient condition for the color confinement is the infrared vanishing of the gluon propagator\cite{Gr}. A sufficient condition for the color confinement based on the Lagrangian field theory satisfying the Becchi-Rouet-Stora-Tyutin(BRST) symmetry was also proposed by Kugo and Ojima\cite{KO}. Zwanziger showed that the uniqueness of the gauge field can be achieved by restricting the gauge field to the fundamental modular region and showed a horizon condition, which can be checked by the lattice simulation\cite{Zw,Zw1}. Investigations of the color-confinement criterion in Landau gauge as well as in the Curci-Ferrari gauge based on the Dyson-Schwinger approach are referred in ref. \cite{AvS}.

Lattice Landau gauge QCD simulation is a valuable tool for analyzing confinement and chiral symmetry breaking from the first principle. In the simulation of
quenched lattice Landau gauge of $\beta=$6, 6.4 and 6.45 with lattice volume $24^4, 32^4, 48^4$ and $56^4$, presence of infrared fixed point of $\alpha_0\sim 2.5$ was suggested and the tendency of increasing Kugo-Ojima parameter $c$ as the
continuum limit is approached was observed\cite{FN03,FN04}. The parameter $c$ which is expected to be 1 for the proof of the confinement remained about 0.8. The lattice Landau gauge QCD suffers from gauge non-uniqueness problem i.e. Gribov copy problem\cite{Gr,scgt} and in large lattice we observed exceptional samples which possess axes
along which the reflection positivity of the 1-d Fourier transform is violated
 and whose Kugo-Ojima parameters are close to 1.  Most of the 1-d Fourier transforms of gauge configurations in quenched simulation violate rotational symmetry, but coupling to fermions recovers the symmetry. Besides this feature, a certain light meson propagator in quenched theory exhibits chiral loop artefacts\cite{bdeit,auog}, and thus the unquenched simulation results could be qualitatively different from those of quenched simulation.

Recently, the MILC collaboration claimed that lattice QCD with three flavors agrees with variety of quantities with both light (u, d and s) and heavy (c or b) quarks with errors of 2-3\%, whereas quenched QCD has errors as large as 15-20\%\cite{got}.
The Kogut-Susskind(KS) fermion contains artificial flavor degrees of freedom and MILC collaboration eliminated  these degrees of freedom by taking the 4th root of the fermion determinant. This procedure can be justified when the flavor (taste) symmetry is preserved, but in the usual KS fermion approach it is violated in square order of the lattice spacing.  The Asqtad action used by the MILC collaboration has an advantage that the taste violation would be reduced when the lattice spacing $a$ is small\cite{milc1}. We need to check this by measuring the running coupling and compare with the results of other fermions such as Wilson fermions, where no taste problem exists. 

We investigated gauge configurations of unquenched simulation in the data base i.e. JLQCD\cite{jlqcd}, CP-PACS\cite{cppacs}, MILC\cite{milc} and Columbia University (CU)\cite{cu,kil}. JLQCD and CP-PACS use Wilson fermions. The former is based on O(a) improved Wilson action with a non-perturbatively defined clover coefficient $c_{SW}$, and the latter is based on the Iwasaki-improved gauge action with the tadpole improved clover coefficient $c_{SW}$.  MILC and CU use KS fermions. MILC is based on the Asqtad improved action i.e. an extension of the L\"uscher-Weisz improved gauge action\cite{Weis,LW} and tadpole-improved fermion action.  CU is based on the old standard Wilson action. 

In chiral perturbation theory, the length scale $L=V^{1/4}$,  the pion mass $m_\pi$, infinite-volume pion decay constant $F$, quark condensate $\Sigma$ and the effective cutoff $\Lambda\simeq 4\pi F$ characterize the system. In order that particles fits well inside the box one requires the Compton wavelength of the pion $1/m_\pi \ll L$. On the other hand, near the chiral limit extrapolation to $1/m_\pi \gg L$ is required so that the collective Goldstone boson can be properly taken into account\cite{ddhj}. In the case of the Wilson fermion, there appears
a problem to bridge the two regions due to appearance of a parity- and isospin-violating Aoki phase, and improvement by the twisted mass fermion etc. is proposed\cite{ab}. Whether KS fermion suffers from the same problem is discussed by several
authors\cite{aubin}. Thus it is important to clarify the infrared features of Wilson fermions and KS fermions on the lattice.

The gauge configurations that we investigate are summarized in Table\ref{datab}. 

\begin{table}[htb]
\beforetab
\begin{tabular}{c c c c c c c c c}
\firsthline
   &$\beta$ &$K_{sea}$ & $am^{VWI}_{ud}/am^{VWI}_{s}$& $N_f$& $1/a$(GeV)&$L_s$ & $L_t$ &$a L_s$(fm)\\
\midhline
JLQCD &5.2&  0.1340  & 0.134 & 2 & 2.221 & 20 & 48&1.78\\
      &5.2&  0.1355 & 0.093  & 2 & 2.221 & 20 & 48&1.78\\
\midhline
CP-PACS & 2.1&  0.1357  & 0.087 & 2 & 1.834 & 24 & 48&2.58\\
        & 2.1 & 0.1382  & 0.020 & 2 & 1.834 & 24 & 48&2.58\\
\midhline
CU &5.415&      & 0.025 & 2& 1.140 & 16 & 32&2.77\\
   &5.7&      & 0.010 & 2& 2.1 & 16 & 32 &1.50\\
\midhline
MILC$_c$ &6.83($\beta_{imp}$)&      & 0.040/0.050& 2+1 & 1.64 & 20 &64&2.41\\
       &6.76($\beta_{imp}$)&      & 0.007/0.050& 2+1 & 1.64 & 20 &64&2.41\\
\midhline
MILC$_f$ &7.11($\beta_{imp}$)&      & 0.0124/0.031& 2+1 & 2.19 &28 & 96&2.52\\
       &7.09($\beta_{imp}$)&      & 0.0062/0.031& 2+1 & 2.19 &28 & 96&2.52\\
\lasthline
\end{tabular}
\aftertab
\captionaftertab[] {$\beta, K_{sea}$ and the sea quark mass $m^{VWI}$(vector Ward identity) and the
inverse lattice spacing $1/a$, lattice size and lattice length(fm). Suffices $c$ and $f$ of MILC correspond to coarse lattice($a$=0.12fm) and fine lattice($a$=0.09fm). $\beta_{imp}=5/3\times \beta$.}\label{datab}
\end{table}

In Sect.2 the unquenched lattice-simulation method is summarized and in Sect.3 numerical results of the gluon propagator, ghost propagator, Kugo-Ojima parameter and the running coupling are given. The conclusion and a discussion are presented in Sect.4. The pQCD formulae\cite{ChRe, Orsay} that are used in the fit of the lattice data are summarized in the Appendix.
 
\section{Unquenched Lattice simulation}

In the present lattice simulation, we adopt the $\log U$ type gauge field definition:
\begin{equation}
U_{x,\mu}=e^{A_{x,\mu}},\ A_{x,\mu}^{\dag}=-A_{x,\mu}.
\end{equation}

The Landau gauge, $\partial A^g=0$ is specified as a stationary point 
of some optimizing functions $F_U(g)$ along gauge orbit\cite{MN, Zw} where $g$ denotes 
gauge transformation, i.e., $\delta F_U(g)=0\ {\rm for\ any\ } \delta g.$

Here $F_U(g)$ for this options is \cite{NF98,FN03}
\begin{equation}
F_U(g)=||A^g||^2=\sum_{x,\mu}{\rm tr}
 \left({{A^g}_{x,\mu}}^{\dag}A^g_{x,\mu}\right),
\end{equation}
Under the infinitesimal gauge transformation
$g^{-1}\delta g=\epsilon$, its variation reads for this defintion 
\[
\Delta F_U(g)=-2\langle \partial A^g|\epsilon\rangle+
\langle \epsilon|-\partial { D(U^g)}|\epsilon\rangle+\cdots,
\]
where the covariant derivativative $D_{\mu}(U)$  reads 
\begin{equation}
D_{\mu}(U_{x,\mu})\phi=S(U_{x,\mu})\partial_\mu \phi+[A_{x,\mu},\bar \phi]
\end{equation}
Here 
\begin{equation}
\partial_\mu \phi=\phi(x+\mu)-\phi(x), {\rm and}
\bar \phi=\dis{\phi(x+\mu)+\phi(x)\over 2} \nonumber
\end{equation}
and the definition of operation $S(U_{x,\mu})B_{x,\mu}$ is given as
\begin{equation}
S(U_{x,\mu})B_{x,\mu}=T({\cal A}_{x,\mu})B_{x,\mu}
\end{equation}
where ${\cal A}_{x,\mu}=adj_{A_{x,\mu}}=[A_{x,\mu},\cdot]$, 
and $T(x)=\dis{{x/2\over {\rm th}(x/2)}}$.

The gauge fixing of the unquenched configuration is essentially the same as that of the quenched configuration. The convergence condition for the conjugate gradient method is less than 5\% in the $L_2$ norm.
We measure the gluon propagator, the ghost propagator, the running coupling and the Kugo-Ojima parameter. Fermions will affect the running coupling through quark condensates, if they exist and will indirectly affect the Kugo-Ojima parameter.

\section{Numerical results}
In this section we show the lattice results and compare the data with the continuum theory based on the effective charge method\cite{grun, chet}. In this method, the Green function $G$ that depends on the scheme and the scale $\mu$ is a solution of the renormalization group equation
\begin{equation}
\mu^2\frac{1}{d\mu^2}G(h,\mu)=\gamma(h)G(h,\mu)
\end{equation}
and the scheme and scale invariant $\hat G$ is expressed as
\begin{equation}
\hat G=G(h,\mu)/f(h)
\end{equation}
where 
\begin{equation}
f(h)=exp(\int^h \frac{dx}{x}\frac{\gamma(x)}{\beta(x)}).\nonumber
\end{equation}
The function $\gamma(x)$ is a function of the coupling constant and not a function of the scale.@In the $\widetilde{MOM}$ scheme, $\mu$ should be chosen such that the $G(h,\mu)$ can be factorized into the scale-dependent part and the cut-off-dependent part. We have chosen $\mu=1.97$GeV which is the inverse lattice spacing of the quenched $\beta=6.0$ configuration and expressed propagators in terms of the effective coupling constant $h$\cite{FN03}. The effective coupling constant is expanded by a parameter $y$  which is a solution of 
\begin{equation}
\frac{1}{y}=\beta_0\log(\mu^2/\Lambda_{\widetilde{MOM}}^2)-\frac{\beta_1}{\beta_0}\log(\beta_0 y).
\end{equation}
where $\Lambda_{\widetilde{MOM}}$ is the cut-off in the $\widetilde{MOM}$ scheme, $\beta_0$ and $\beta_1$ are scheme independent constants of the $\beta$ function.
A similar analysis using the principle of minimal sensitivity\cite{PMS}(PMS) was performed in \cite{vanacol}, in which $\mu$ is not fixed and the optimal parameter $y$ is searched for each $\mu$. The fit of the ghost propagator in the PMS was not successful, and we adopted the $\widetilde{MOM}$ scheme. Details of the effective charge method are summarized in the Appendix.

\subsection{The gluon propagator}
In the $\log-U$ definition 
we express the gauge field defined at the midpoint of the link as 
\begin{equation}
U_{x,\mu}=\exp A_{x,\mu}=\exp [\frac{\lambda^a}{ 2}A_\mu^a(x+\frac{1}{2}e^\mu)]
\end{equation}
where  $\lambda^a$ is the Gell-Mann's $\lambda$ normalized as tr$\displaystyle \frac{\lambda^{a\dagger}}{2}\frac{\lambda^b}{2}=\frac{\delta^{ab}}{2}$.

This definition is consistent with the continuum theory\cite{FadSlav}, 
but differ from the usual lattice convention\cite{luscher}
\begin{equation}
U_\mu^a(x)=\exp [\lambda^a A_\mu^a].
\end{equation}
where the lattice spacing is taken as 1 and the coupling constant is included in $A_\mu^a$.

The gluon propagator $D_A(q^2)$ in the Landau gauge is defined as
\begin{eqnarray}\label{scalarfn}
D_{A,\mu\nu}(q)&=&\frac{2}{N_c^2-1}{\rm tr}\langle \tilde A_\mu(q)\tilde A_\nu^\dagger(q) \rangle \nonumber\\
&=&(\delta_{\mu\nu}-{q_\mu q_\nu\over q^2})D_A(q^2),
\end{eqnarray}
where $N_c$ is the number of color and 
the Fourier transform of the gauge field is 
\begin{equation}
\tilde A_\mu(q)=\frac{1}{\sqrt V}\sum_x e^{-iq\cdot (x+\frac{1}{2}e^\mu)}A_{x,\mu}.
\end{equation}

The summation over $\mu$  gives
\begin{equation}
\frac{2}{N_c^2-1}\sum_\mu{\rm tr} \langle \tilde A_\mu(q) \tilde A_\mu^\dagger(q)\rangle =(d-1)D_A(q^2) 
\end{equation}
The midpoint definition is crucial in the definition of the momentum of the L\"uscher-Weisz's improved action. In the practical calculation, however, one is allowed to perform the simple numerical Fourier transform 
\begin{eqnarray}
\hat A_\mu(q)&=&\frac{1}{\sqrt V}\sum_x e^{-iq\cdot x}A_{x,\mu}\nonumber\\
&=&e^{iq\cdot \frac{1}{2}e^\mu}\tilde A_\mu(q)
\end{eqnarray}
and evaluate
\begin{equation}
\frac{2}{N_c^2-1}\sum_\mu{\rm tr} \langle \hat A_\mu(q) \hat A_\mu^\dagger(q)\rangle =(d-1)D_A(q^2) 
\end{equation}
which yields trivially the same $D_A(q^2)$ as that obtained by the $\tilde A_\mu(q)$.

In passing, we remark that the Ansatz of the gauge transformation adopted by \cite{LW}
\begin{equation}
\tilde U_\mu(x)=e^{\epsilon X_\mu(x)}U_\mu(x)
\end{equation}
where in the classical continuum limit $\displaystyle X_\mu(x)=\frac{1}{2}a^3\sum_\nu D_\nu F_{\mu\nu}(x)$ yields transformation
\begin{equation}
A_\mu\to A_\mu+\frac{\epsilon}{2}a^2\sum_\nu (\partial_\nu F_{\mu\nu}(x)+[A_\nu,F_{\mu\nu}(x)])+o(a^4)
\end{equation}
In our Ansatz
\begin{equation}
\tilde U_\mu(x)=g^\dagger_x U_\mu(x) g_{x+\mu}
\end{equation}
where $\displaystyle g^\dagger_x=\exp[\frac{\epsilon}{2} a^2\sum_\nu F_{\mu\nu}^\dagger(x)]$ 
and $\displaystyle g_{x+\mu}=\exp[\frac{\epsilon}{2}a^2\sum_\nu F_{\mu\nu}(x+\mu)]$, yields the same expressioin of $o(a^2)$ in the Landau gauge. In the derivation of the L\"uscher-Weisz improved action, the definition
 $U_{\mu R}(x)=\exp a A_\mu^i R^i(x)$  was used\cite{Weis}, where $R$ is an irreducible representation of SU(N).

In the data analysis of $D_{A,\mu\nu}(q)$, we usually adopt $q$ diagonal in the momentum lattice, which is called cylinder cut.
In the case of different lattice spacial length $N_s$ and time length $N_t$, we define cylinder cut as $qa$ around the diagonal $[\tilde q_1,\tilde q_2,\tilde q_3,\tilde q_4]=[\tilde q,\tilde q,\tilde q,(N_t\cdot \tilde q/N_s)]$ where $(N_t\cdot \tilde q /N_s)$ is the closest integer to the quotient of $N_t\cdot \tilde q/N_s$.

When the improved action is adopted, the lattice gluon momentum is defined as\cite{Weis, bhlpw}
$q=\sqrt{{\tilde q}^2+{\tilde q}^4/3}/a$ where
\begin{equation}
{\tilde q}^2 = 4 (\sum_{i=1}^3\sin^2({\tilde q}_i \pi/N_s) + \sin^2({\tilde q}_4 \pi/N_t))
\end{equation}
\begin{equation}
{\tilde q}^4 = 4 (\sum_{i=1}^3\sin^4({\tilde q}_i \pi/N_s) + \sin^4({\tilde q}_4 \pi/N_t))
\end{equation}
The correction factor $\displaystyle \sqrt{1 + \frac{{\tilde q}^4}{3{\tilde q}^2}}$ is about 1.15 in the highest momentum point of cylinder cut.

\begin{figure}[hbt]
\epsfig{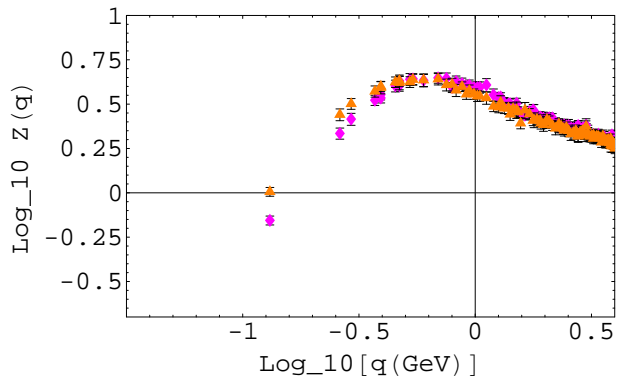}
\caption[]{The $\log$ of the gluon dressing function of CP-PACS $K_{sea}=0.1357$(diamonds)(50 samples) and $0.1382$(triangles)(50samples).
}\label{gldcppacs}
\end{figure}

\begin{figure}[htb]
\epsfig{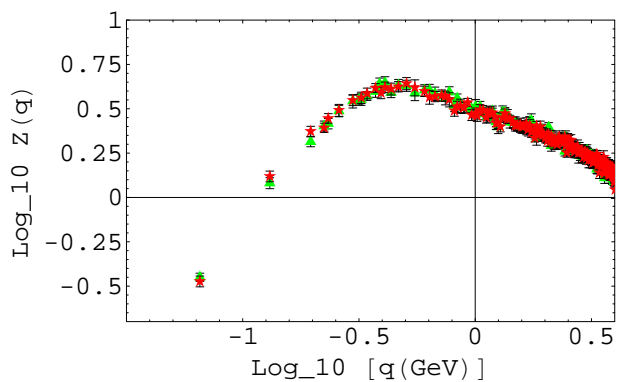}
\caption[]{The $\log$ of the gluon dressing function of MILC $\beta_{imp}=7.09$(stars)(50 samples) and $7.11$(triangles)(50samples).
}\label{gldlmilc}
\end{figure}

In Fig. \ref{gldcppacs}, the logarithm of the gluon dressing function as a function of
 $\log_{10} q(GeV)$ of the Wilson fermion $K_{sea}=0.1357$ and $K_{sea}=0.1382$ are shown. 
The corresponding data of the KS fermion $\beta_{imp}=7.11$ and 7.09 are shown in Fig. \ref{gldlmilc}. 
In the infrared region the dressing function of light Wilson fermion mass ($K_{sea}=0.1382$) is enhanced as compared to that of the heavy Wilson fermion mass ($K_{sea}=0.1357$), while that of heavy KS fermion mass ($\beta_{imp}=7.11$) and that of light KS fermion mass ($\beta_{imp}=7.09$) are almost identical.


\begin{figure}[htb]
\epsfig{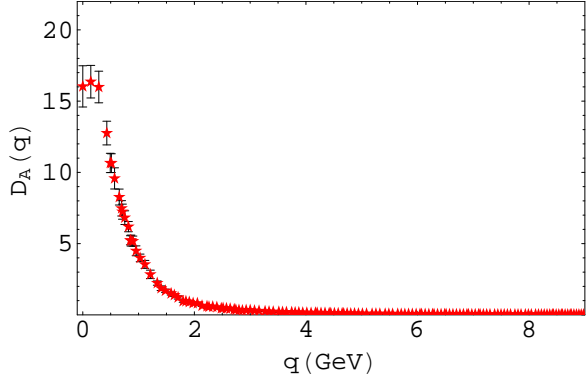}
\caption[]{The gluon propagator of MILC$_f$ $\beta_{imp}=7.09$.
}\label{gl709}
\end{figure}

\begin{figure}[htb]
\epsfig{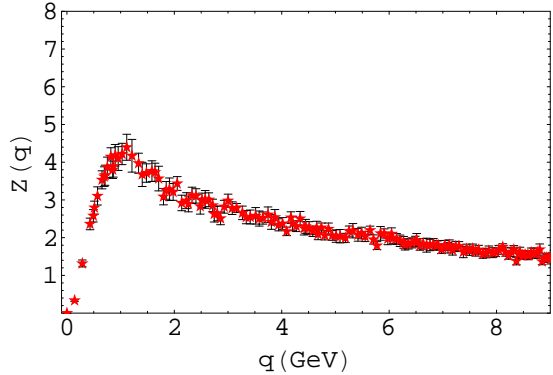}
\caption[]{Same as Fig. \ref{gl709} but the gluon dressing function. 
}\label{gld709}
\end{figure}

In Figs. \ref{gl709} and \ref{gld709}, the gluon propagator and the gluon dressing function of MILC fine lattice (MILC$_f$, $\beta_{imp}=7.09$) in cylinder cut are shown. Data of MILC coarse lattice (MILC$_c$) are consistent with those of \cite{bhlpw}.  In \cite{bhlpw}, the gluon propagator is normalized as $1/q^2$ at $q=4$GeV and the data are about factor 2 smaller than ours.


\begin{figure}[htb]
\epsfig{file=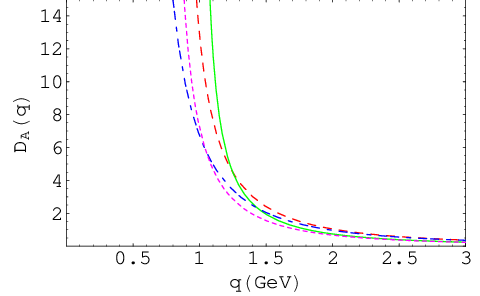,scale=1.0}
\caption[]{The gluon propagator of pQCD 3-loop $N_f=0$(short-dashed),$N_f=3$(dash-dotted).
4-loop $N_f=0$(solid),$N_f=3$(dashed).
}\label{glp}
\end{figure}

\begin{figure}[htb]
\epsfig{file=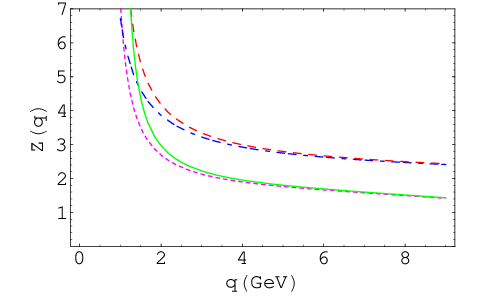,scale=1.0}
\caption[]{Same as Fig. \ref{glp} but the gluon dressing function. 
}\label{gld}
\end{figure}

The pQCD formulae of the gluon dressing function are given in the Appendix.
We observe that the gluon propagator of pQCD in quenched approximation is larger than that of the unquenched one and the result of the four-loop calculation is larger than that of the three-loop calculation.(Fig. \ref{glp}) The corresponding data of gluon dressing function are
shown in Fig. \ref{gld}. In these plots we used $\Lambda_{\overline{\rm MS}}=0.237$GeV, $\mu=1.97$GeV, $y=0.0222703$ and $\lambda=17.85$ to fit $Z(9.5$GeV)=1.3107 \cite{Orsay2} obtained in the quenched Landau gauge simulation. The definition of the effective coupling strength $y$ and the strength $\lambda$ are given in the Appendix.


\begin{figure}[htb]
\epsfig{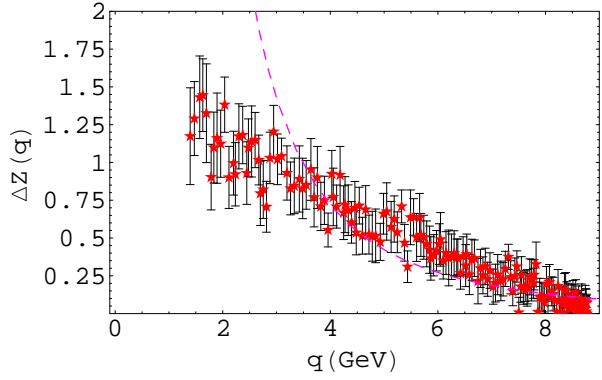}
\caption[]{The difference $\Delta Z(q^2)= Z_{latt}(q^2,\mu^2)-Z_{pert}(q^2,\mu^2)$ as a function of $q($GeV). The dashed line is a fit by condensates of dimesion-2. 
}\label{dglmilc}
\end{figure}

The difference of $Z(q)$ between lattice data and pQCD results would yield
information on the gluon condensates\cite{bllmpr}. We performed $\chi^2$ fit of the difference of the gluon dressing function of MILC$_f$ $\beta_{imp}=7.09$ and pQCD four-loop $N_f=3$ result. We parametrize
\begin{equation}
Z_{latt}(q^2,\mu^2)=Z_{pQCD}(q^2,\mu^2)(1+\frac{\tilde c_1}{q^2})+d
\end{equation}
where $\mu=8.77$GeV and the data of $q>3$GeV region fit by searching $\tilde c_1$ and $d$.
Fig.\ref{dglmilc} is the fitting result using the pQCD result for $\mu=8.78$GeV, $y=0.0148488$ which gives $\tilde c_1=7.39$GeV$^2$, $d=-0.024$, $\chi^2/d.o.f=1.10$ . The fit using $\mu=1.97$GeV gives
$\tilde c_1=7.04$GeV$^2$, $d=-0.017$ and $\chi^2/d.o.f=1.14$. 

\subsection{The ghost propagator}
The ghost propagator is defined by  the Fourier transform(FT) of the  expectation value of the inverse Faddeev-Popov(FP) operator $\cal  M$
\begin{eqnarray}
FT[D_G^{ab}(x,y)]&=&FT\langle {\rm tr} ( \Lambda^{a \dagger} \{({\cal  M}[U])^{-1}\}_{xy}
\Lambda^b  \rangle,\nonumber\\
&=&\delta^{ab}D_G(q^2).
\end{eqnarray}
where antihermitian $\Lambda^a$ is normalized as ${\rm tr} \Lambda^{a\dagger} \Lambda^b=\delta^{ab}$.

The ghost dressing function $G(q^2)$ is defined as
\begin{equation}\label{dgg}
D_G(q^2)=\frac{G(q^2)}{q^2}.
\end{equation}

\begin{figure}[htb]
\epsfig{file=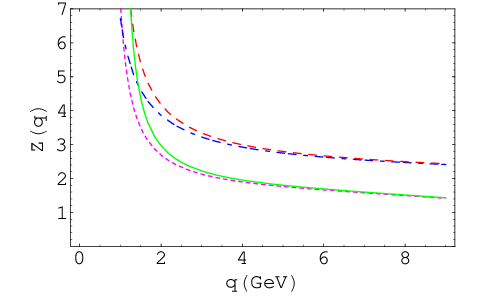,scale=1.0}
\caption[]{The ghost propagator of pQCD 3-loop $N_f=0$(short-dashed), $N_f=3$(dash-dotted).
4-loop $N_f=0$(solid),$N_f=3$(dashed).
}\label{ghp}
\end{figure}

\begin{figure}[htb]
\epsfig{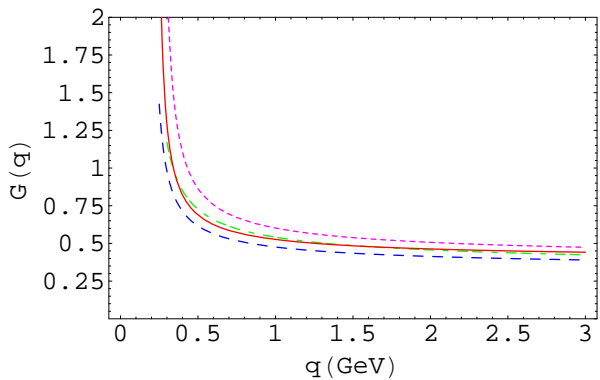}
\caption[]{Same as Fig. \ref{ghp} but the ghost dressing function. 
}\label{ghd}
\end{figure}

The pQCD formula of the ghost dressing function is given in the Appendix. We observe that the ghost propagator of pQCD in quenched approximation is larger than that of the unquenched one, and the result of the four-loop calculation is smaller than that of the three-loop calculation.(Fig. \ref{ghp}) The corresponding data of the ghost dressing function are shown in Fig. \ref{ghd}.  

In these plots we used $\Lambda_{\overline{\rm MS}}=0.237$GeV, $\mu=1.97$GeV, $y=0.0158465$ and $\lambda_G=1$. 

\begin{figure}[htb]
\epsfig{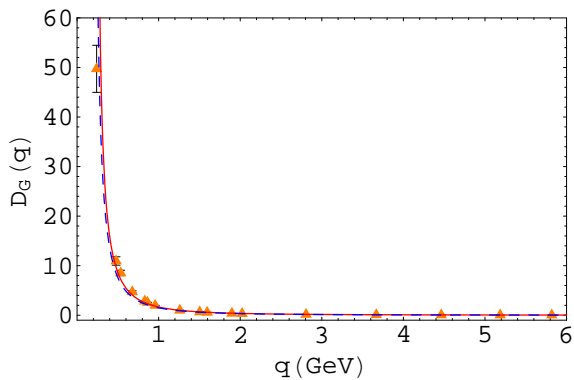}
\caption[]{The ghost propagator as a function of $q$(GeV) of CP-PACS $K_{sea}=0.1382$(50 samples). Solid curve is the 3-loop $N_f=0$ pQCD result and dashed curve is the 4-loop $N_f=2$ pQCD result ($\lambda_G=3.22, y=0.0158465$). 
}\label{ghcppacs}
\end{figure}

\begin{figure}[htb]
\epsfig{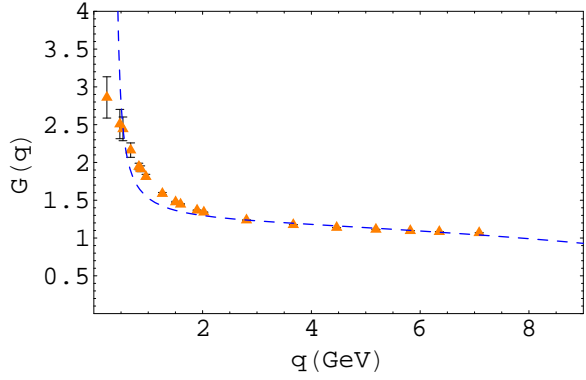}
\caption[]{The ghost dressing function as a function of  $q$(GeV) of CP-PACS $K_{sea}=0.1382$(50 samples).  Dashed line is the 4-loop $N_f=2$ pQCD result ($\lambda_G=3.01, y=0.0246100$). 
}\label{ghdcppacs}
\end{figure}

The ghost propagator of CP-PACS $K_{sea}=0.1382$ is shown in Figs. \ref{ghcppacs}. The solid curve in Fig. \ref{ghcppacs} is the pQCD result in the three-loop $N_f=0$\cite{FN03}, using the scale parameter $\mu=1.97$GeV and $\lambda_G=3.22$ that fit the quenched data. The dashed line is the pQCD result in the four-loop $N_f=2$ using the same $\lambda_G$ and $y$. In the region $2<q<7$GeV the pQCD result of $N_f=2$ is about 10\% smaller than that of $N_f=0$ and parameters $\lambda_G=3.22, y=0.0158465$ give qualitatively good agreement. In order to get better agreement of CP-PACS, we change $y=0.024610$ as the solution of $N_f=2$ and perform $\chi^2$ fit of $\lambda_G$ using the data of ghost dressing function in $q>2.8$GeV region. The fit with $\lambda_G=3.01$ is shown in Fig.\ref{ghdcppacs}


\begin{figure}[htb]
\epsfig{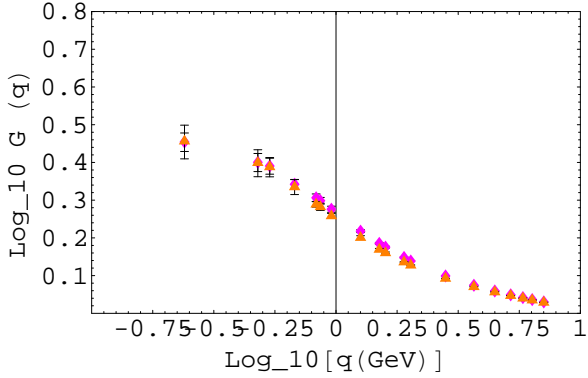}
\caption[]{The $\log_{10} G(q)$ as a function of  $\log_{10} q$(GeV) of CP-PACS $K_{sea}=0.1357$(diamonds)(50 samples) and $K_{sea}=0.1382$(triangles)(50samples).
}\label{ghdlcppcs}
\end{figure}

\begin{figure}[htb]
\epsfig{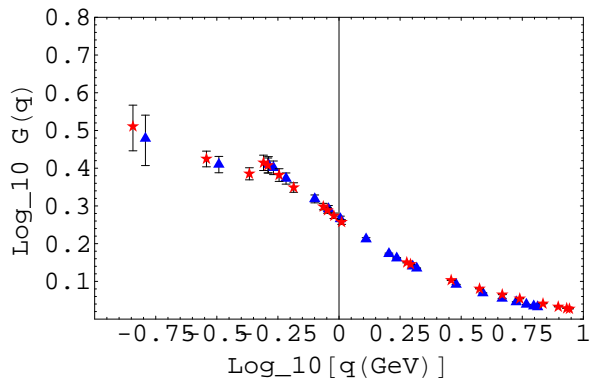}
\caption[]{The $\log_{10} G(q)$ as a function of  $\log_{10} q(GeV)$  of MILC$_f$ $\beta_{imp}=7.09$(stars)(6 samples) and MILC$_c$ $\beta_{imp}=6.76$(triangles)(50 samples). 
}\label{ghdlmilc}
\end{figure}

The logarithm of the ghost dressing function as a function of $\log_{10} q(GeV)$ of the Wilson fermion (CP-PACS) and the KS fermion (MILC) are shown in Figs.\ref{ghdlcppcs} and \ref{ghdlmilc}, respectively.
In the case of CP-PACS, the ghost dressing functions are almost independent of the $K_{sea}$ and the exponent $\alpha_G$ in the region $q>0.4$GeV is 0.22(5), and in the case of the MILC fine lattices it is 0.24(5) in $\beta_{imp}=7.09$ and 0.23(5) in $\beta_{imp}=7.11$.  The exponent $\alpha_G$ of the quenched, the unquenched Wilson and the KS fermions are almost the same. 
 In view of the discontinuity in the slope of the ghost propagator in the infrared region, we ignore the lowest few points in the evaluation of the running coupling. In the asymptotic region ($\log_{10} q(GeV)\sim 1$) the ghost dressing function of MILC$_c$, MILC$_f$ and CP-PACS converge  as shown in Figs.\ref{ghdlcppcs} and \ref{ghdlmilc}.


We performed $\chi^2$ fit of $\lambda_G$ for ghost dressing function of MILC$_f$.  Using $\mu=1.97$GeV and $N_f=3$, we obtained $y=0.026775$ but with this parameter the fit was not better than that using $y=0.024610$ i.e. the solution of $N_f=2$. We obtained $\lambda_G=3.258, y=0.024610$ for $\beta_{imp}=7.09$ data and  $\lambda_G=3.237, y=0.024610$ for $\beta_{imp}=7.11$ data.  Since the sea-quark mass $m_s$ is large, the better fit using pQCD with $N_f=2$ is not unexpected. It is remarkable that $\lambda_G$ of MILC$_f$ are close to that of quenched simulations, i.e. $\lambda_G=3.22$.

\subsection{The Kugo-Ojima parameter} 

The Kugo-Ojima parameter is defined by the two point function of the
covariant derivative of the ghost and the commutator of the antighost and gauge field
\begin{eqnarray}
&&\left(\delta_{\mu\nu}-{q_\mu q_\nu\over q^2}\right)u^{ab}(q^2)\nonumber\\
&&={1\over V}
\sum_{x,y} e^{-ip(x-y)}\left\langle {\rm tr}\left({\Lambda^a}^{\dag}
D_\mu \displaystyle{1\over -\partial D}[A_\nu,\Lambda^b] \right)_{xy}\right\rangle.\nonumber\\
\end{eqnarray}
The confinement criterion is that the parameter $c$ defined as $u^{ab}(0)=-\delta^{ab}c$ becomes 1. The parameter is related to the renormalization factors which are defined in the next section as\cite{Kugo}
\begin{equation}
1-c=\frac{Z_1}{Z_3}=\frac{\tilde Z_1}{\tilde Z_3}
\end{equation}
If the finiteness of $\tilde Z_1$ is proved, divergence of $\tilde Z_3$ is a
sufficient condition. If $Z_3$ vanishes in the infrared, $Z_1$ should have higher order 0.

From the investigation of the Gribov problem, 
Zwanziger proposed the horizon function\cite{Zw,Zw1}
\begin{eqnarray}\label{horizon}
&&\frac{1}{V}\sum_{x,y} e^{-iq(x-y)} \left \langle {\rm tr}\left({\Lambda^a}^{\dag}
D_\mu \displaystyle{1\over -\partial D}(-D_\nu)\Lambda^b\right)_{xy}\right
\rangle\nonumber\\
&&=G_{\mu\nu}(q)\delta^{ab}\nonumber\\
&&=\left(\displaystyle{e\over d}\right)\displaystyle{q_\mu q_\nu\over q^2}\delta^{ab}
-\left(\delta_{\mu\nu}-\displaystyle{q_\mu q_\nu\over q^2}
\right)u^{ab},
\end{eqnarray}
where $\displaystyle e=\left\langle\sum_{x,\mu}{\rm tr}(\Lambda^{a\dag} 
S(U_{x,\mu})\Lambda^a)\right\rangle/\{(N_c^2-1)V\}$, and $N_c=3$ for SU(3).
The horizon condition reads $\displaystyle \lim_{q\to 0}G_{\mu\mu}(q)-e=0$,
and the left-hand side of the condition is 
$
\left(\displaystyle{e\over d}\right)+(d-1)c-e=(d-1)h
$
where $h=c-\dis{e\over d}$ and dimension $d=4$, and it follows that 
$h=0$ implies the validity of the horizon condition, and thus the horizon condition coincides
with Kugo-Ojima criterion provided the covariant derivative approaches
the naive continuum limit, i.e., $e/d=1$.

Therefore the infrared behavior of the gluon propagator and the ghost propagator, as well as direct measurement of the Kugo-Ojima parameter are important for understanding the color-confinement mechanism.

The Kugo-Ojima parameter $c$ of the unquenched simulation depends on the direction of the polarization due to asymmetry of the lattice. When the polarization is in the spacial directions, $c$ is consistent with 1 in most unquenched simulations. (See Table \ref{kugo})

\begin{table}[htb]
\beforetab
\begin{tabular}{c c c c c c c}
\firsthline
&$K_{sea}$ or $\beta$ & $c_x$     & $c_t$    &$c$ &  $e/d$        &    $h$     \\
\midhline
JLQCD&$K_{sea}=$0.1340 & 0.89(9)  & 0.72(4) &0.85(11) & 0.9296(2) & -0.08(11)  \\
&$K_{sea}=$0.1355 & 1.01(22) & 0.67(5) &0.92(24) & 0.9340(1) & -0.01(24)  \\
\midhline
CP-PACS&$K_{sea}=$0.1357 & 0.86(6)  & 0.76(4) &0.84(7) & 0.9388(1) & -0.10(6)  \\
&$K_{sea}=$0.1382 & 0.89(9) & 0.72(4) &0.85(11) & 0.9409(1) & -0.05(9)  \\
\midhline
CU&$\beta=$5.415 & 0.84(7)  & 0.74(4) &0.81(8) & 0.9242(3) & -0.11(8)  \\
&$\beta=$5.7 & 0.95(26)  & 0.58(6) &0.86(28) & 0.9414(2) & -0.08(28)  \\
\midhline
MILC$_c$&$\beta=$6.76 & 1.04(11)  & 0.74(3) &0.97(16) & 0.9325(1) & 0.03(16)  \\
&$\beta=$6.83 & 0.99(14)  & 0.75(3) &0.93(16) & 0.9339(1) &  -0.00(16) \\
\midhline
MILC$_f$&$\beta=$7.09 & 1.06(13)  & 0.76(3) &0.99(17) & 0.9409(1) & 0.04(17)  \\
  &$\beta=$7.11 &1.05(13)   &  0.76(3)& 0.98(17) & 0.9412(1) &  0.04(17) \\
\lasthline
\end{tabular}
\aftertab
\captionaftertab[]{The Kugo-Ojima parameter for the polarization along the spacial directions $c_x$ and that along the time direction $c_t$ and the average $c$, trace divided by the dimension $e/d$, horizon function deviation $h$ of the unquenched Wilson fermion(JLQCD, CP-PACS), and KS fermion (MILC$_c$,CU,MILC$_f$).  The  $\log U$ definition of the gauge field is adopted. }\label{kugo}
\end{table}

\subsection{The running coupling}
The calculation of the running coupling in the $\widetilde{MOM}$ scheme using the gluon dressing function and the ghost dressing function is discussed in \cite{vSHA}.
We parametrize the gluon dressing function as
\begin{equation}
Z_R(q^2,\mu^2)=Z_3^{-1}(\beta,\mu^2)Z(q^2)\label{glsc}
\end{equation}
and the ghost dressing function as
\begin{equation}
G_R(q^2,\mu^2)=\tilde Z_3^{-1}(\mu^2)G(q^2)\label{ghsc}
\end{equation}
with the renormalization conditions
\begin{equation}
Z_R(\mu^2,\mu^2)=G_R(\mu^2,\mu^2)=1
\end{equation}
where $\mu$ is an accessible scale of the lattice simulation\cite{bclm1}. We choose $\mu\sim 6$GeV . Infrared properties of $Z(q^2)$ and $G(q^2)$ are defined as $Z(q^2)\propto (qa)^{-2\alpha_A}$ and $G(q^2)\propto (qa)^{-2\alpha_G}$. 

We define the vertex renormalization factor $\tilde Z_1$ as
\begin{eqnarray}\label{alphadef}
&&\alpha_R(\mu^2)Z_R(q^2,\mu^2){G_R}(q^2,\mu^2)^2\nonumber\\
&&=\frac{\alpha_0(\Lambda_{UV})}{{\tilde Z_1}^2(\beta,\mu)}\times Z(q^2){G}(q^2)^2
\end{eqnarray}
where the subscript  $R$ means "renormalized", and
\begin{equation}
\alpha_R(q^2)=\alpha_R(\mu^2)Z_R(q^2,\mu^2){G_R}(q^2,\mu^2)^2
\end{equation}

The MILC collaboration adopted the tadpole improvement in the generation of the gauge configuration. In \cite{bclm1} the gluon propagator is calculated as
\begin{equation}
\alpha_0(\Lambda_{UV}) a^2 {\tilde D}_A(x,y)=D_A(x,y)/u_{0,P}^2
\end{equation}
where ${\tilde D}_A(x,y)$ corresponds to the propagator in terms of the link matrices, and the ghost propagator as
\begin{equation}
\alpha_0(\Lambda_{UV}) a^2 {\tilde D}_G(x,y)=D_G(x,y) u_{0,P}
\end{equation}
where $\displaystyle u_{0,P}=\langle P \rangle^{1/4}$.

The plaquette value $\langle P \rangle$ of MILC$_c$ is smaller than that of MILC$_f$ by a few percent and the correction in the Fig.\ref{ghdlmilc} by this renormalization is negligible. In the running coupling, the tadpole factors cancel out\cite{bclm1}.
We modify the notation of eq.(\ref{alphadef}) and measure running coupling $\alpha_s(q)$ as\cite{FN05}
\begin{equation}\label{alpha}
\alpha_s(q)=\frac{g_0^2}{4\pi}\frac{Z(q^2){ G({\tilde q}^2/a^2)}^2}{{\tilde Z_1}^2}\sim \alpha_s(\Lambda_{UV}) {\tilde q}^{-2(\alpha_D+2\alpha_G)},
\end{equation}
where $q=\sqrt{{\tilde q}^2+{\tilde q}^4(\delta/3)}/a$. ($\delta=0$ for ordinary action and $\delta=1$ for the improved action.)

In the quenched simulation the Orsay group fitted the lattice data by $\alpha_{s,latt}(\mu)=\alpha_{s,pert}(\mu)+c_1/\mu^2$ with $c_1\sim 0.65$GeV$^2$\cite{Orsay}. In the analysis of unquenched Wilson fermion data\cite{Orsay3,ggmr}, they fitted the lattice data in the form
\begin{equation}
\alpha_{s,latt}(\mu)=\alpha_{s,pert}(\mu)(1+\frac{c_1}{\mu^2})
\end{equation}
with $c_1\sim 2.8(2)$GeV$^2$.

In the quenched simulation we confirmed the correction term\cite{FN04} and we studied
whether the same correction appears in the unquenched configuration.
We define the scale of the running coupling by fitting $\tilde Z_1$ such
that the running coupling at $q\sim 6$GeV agrees with the pQCD results of $N_f=2$(JLQCD,CP-PACS,CU) or $N_f=3$(MILC).  In the case of the improved action, there is an ambiguity due to the mismatch of the momentum of the ghost dressing function $\tilde q/a$ and that of the gluon dressing function $q$. 

\begin{table}[htb]
\beforetab
\begin{tabular}{c c c c c}
 config.  & heavy  & light & $N_f$ &comments\\
\firsthline
JLQCD &0.90(7)&  0.97(7) & 2 & $K_{sea}=0.1340, 0.1355$\\
CP-PACS &1.07(8)&  1.21(10) & 2 & $K_{sea}=0.1357,0.1382$\\
CU &  1.13(10) & 1.19(8) & 2 &$\beta=5.415, 5.7$\\
MILC$_c$ &  1.49(11) & 1.43(10)& 3 &$\beta_{imp}=6.83, 6.76$\\
MILC$_f$ & 1.37(9) & 1.41(12)  & 3 &$\beta_{imp}=7.11, 7.09$\\
\lasthline
\end{tabular}
\aftertab
\captionaftertab[]{The $1/\tilde Z_1^2$ factor of the unquenched SU(3).}\label{z1fac}
\end{table}

The magnitude of the running coupling $\alpha_s(q)$ in the infrared is roughly proportional
to the $1/\tilde Z_1^2$ factor.  

\begin{figure}[htb]
\epsfig{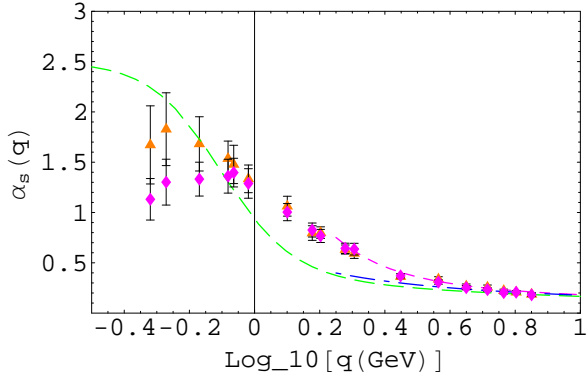}
\caption[]{The running coupling $\alpha_s(q)$ as a function of  $\log_{10}q$(GeV) of CP-PACS $\beta=2.1$, $24^3\times 48$ lattice,  $K_{sea}=0.1357$ (diamonds) and that of 0.1382 (triangles), (25 samples each). The long dashed line is the Dyson-Schwinger fit with $\alpha_0=2.5$, the dash-dotted line is the pQCD result and the short dashed line is the pQCD$\times(1+c/q^2)$ correction. 
}\label{alpcppacs}
\end{figure}

The mass of the sea quark is relatively heavy in JLQCD. There is a deviation from the pQCD in the region $ q <$ 3GeV,  but above 3GeV the
deviation is within statistical errors.
The CP-PACS configuration has a lower sea quark mass and an Iwasaki-improved action is used for the gauge 
action. The running coupling of CP-PACS is shown in Fig. \ref{alpcppacs}.
The absolute value increases as the mass of sea quark decreases and the data of lightest quark mass suggest a maximum of ${\alpha_s} \sim 2-2.5$.

We measured the running coupling of the KS fermion in small $\beta$ (strong coupling region) using the gauge configurations of CU\cite{cu,kil}, and the large $\beta$ using those of MILC. 
In contrast to the Wilson fermion, the absolute
value of the running coupling is close to that of JLQCD $K_{sea}=0.1355$ and does not depend on the mass of the sea quark.

\begin{figure}[htb]
\epsfig{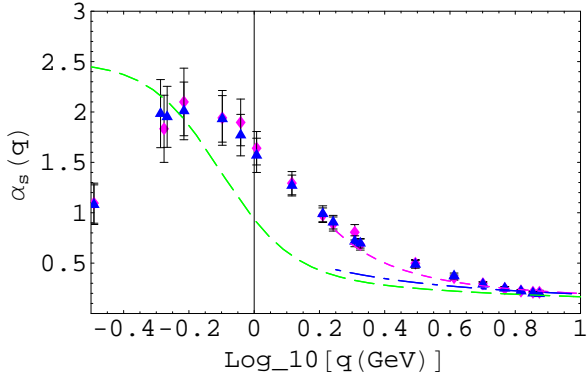}
\caption[]{Same as Fig. \ref{alpcppacs} but the data of MILC$_c$ ($a=0.12$fm) $\beta_{imp}=6.76$(triangles) and 6.83(diamonds), (50 samles each).  }
\label{alpmilcl}
\end{figure}

\begin{figure}[htb]
\epsfig{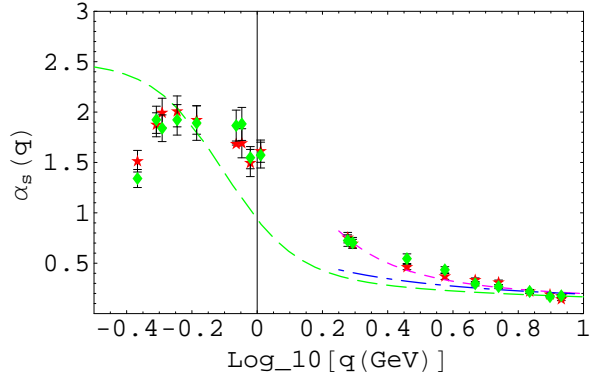}
\caption[]{Same as Fig. \ref{alpcppacs} but the data of MILC$_f$ ($a=0.09$fm) $\beta_{imp}=7.09$(stars) and 7.11(diamonds) (15samples each). 
}\label{alp709_711l}
\end{figure}

The MILC collaboration improved the flavor symmetry violation in the KS fermion by choosing an appropriate improved fermion action which is called Asqtad action. The running coupling of MILC$_c$ and MILC$_f$ are shown in Figs. \ref{alpmilcl}
and  \ref{alp709_711l}, respectively. 

 We observe that the absolute value is consistent with that of the CP-PACS of the smallest sea quark mass. The correction of $c_1/q^2$ with
$c_1$ of the order of 2.8GeV$^2$ observed by the Orsay group in the Wilson fermion exists also in the KS fermion of Asqtad action. 

\begin{figure}[htb]
\epsfig{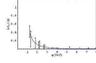}
\caption[]{$\Delta \alpha_s$ of JLQCD $K=0.1355$ as a function of $q$(GeV).
The fitted lines have $c_1=1.12$GeV$^2$ and $c_1=1.19$GeV$^2$, respectively.
}\label{dalp355}
\end{figure}

\begin{figure}[htb]
\epsfig{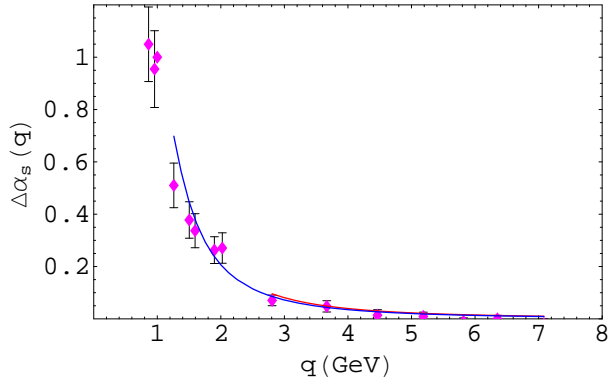}
\caption{$\Delta \alpha_s$ of CP-PACS $K=0.1382$ as a function of $q$(GeV).
The fitted lines have $c_1=2.57$GeV$^2$ and $c_1=2.38$GeV$^2$, respectively.
}\label{dalp382}
\end{figure}

In Figs. \ref{dalp355} and \ref{dalp382} the difference of running coupling of lattice data and pQCD data, $\Delta \alpha_s=\alpha_{s,latt}-\alpha_{s,pert}$ as a function of $q$(GeV) for JLQCD $K_{sea}=0.1355$,  CP-PACS $K_{sea}=0.1382$
 are plotted, respectively. 
  The fitted curves are $\alpha_s(q) c_1/q^2+d$ where the parameters $c_1$ and $d$ are obtained by the $\chi$-square fit using data above $\sim 3$GeV (solid) and data above $\sim 1$GeV (dashed), respectively.  In these configurations and in MILC$_c$ $\beta_{imp}=6.83$, the parameter $c_1$ obtained by the two fits agrees within the statistical errors.  The parameter $c_1\sim 2.4(2)$GeV$^2$ of CP-PACS is 15\%  smaller than the Orsay fit of $K_{sea}=0.15$ Wilson fermion data. The parameter $c_1\sim 1.9(3)$GeV$^2$ of MILC$_c$ $\beta_{imp}=6.83$ is about 2/3 of the Orsay fit.  The $\alpha_s$ of JLQCD have $c_1\sim 1.15(4)$GeV$^2$ but above 3GeV it is consistent with pQCD. 

\begin{figure}[htb]
\epsfig{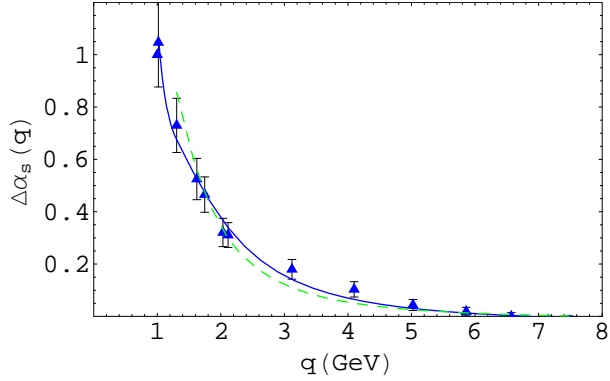}
\caption{$\Delta \alpha_s$ of MILC$_c$ $\beta=6.76$ as a function of $q$(GeV).
The fitted lines have $c_1=6.50$GeV$^2$,$c_2=-11.70$GeV$^4$, $c_3=7.47$GeV$^6$ and d=-0.29 (solid) and  $c_1=4.18$GeV$^2$,$c_2=-2.45$GeV$^4$ and $d=-0.014$ (dashed).
}\label{dalp676}
\end{figure}

\begin{figure}[htb]
\epsfig{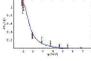}
\caption[]{$\Delta \alpha_s$ of MILC$_f$ $\beta_{imp}$=7.11(diamonds) and 7.09(stars) as a function of $q$(GeV).
The solid line is the fit of $\beta_{imp}=7.11$($c_1=4.27$GeV$^2$, $c_2=-2.28$GeV$^4$)  and dashed line is the fit of $\beta_{imp}=7.09$ ($c_1=4.29$GeV$^2$, $c_2=-2.30$GeV$^4$). 
}\label{dalp709}
\end{figure}

In the case of MILC$_c$ $\beta=6.76$ and MILC$_f$, $\Delta \alpha_s$ cannot be fitted by the factor $\displaystyle \frac{c_1}{q^2}$, therefore we fitted above 1GeV region by the factor $\displaystyle \frac{c_1}{q^2}+\frac{c_2}{q^4}+\frac{c_3}{q^6}$ and the overall shift $d$. In the theory of operator product expansion, $c_2$ is proportional to the gluon condensates $\langle g^2 F_{\mu\nu}^2\rangle$\cite{bllmpr,soro} and/or quark condensates $\langle m\bar q q\rangle$, and $c_6$ is proportional to condensates of dimension 6 like $\langle g^3f^{abc} F^a_{\mu\nu}F^b_{\nu\gamma}F^c_{\gamma\mu}\rangle$. We performed $\chi^2$ fit of $c_{i}$ (i=1,2,3) by either fixing $c_3=0$ or without fixing.
As shown in Table \ref{cndnsts}, we find  
$c_1\sim 4.2(1)$GeV$^2$, $c_2\sim -2.3(2)$GeV$^4$, $c_3=0$ or $c_1= 6.6(1)$GeV$^2$, $c_2= -13(2)$GeV$^4$, $c_3= 8(2)$GeV$^6$ as an average of $\beta_{imp}=$6.76, 7.09 and 7.11 data.  The fit of $\Delta \alpha_s$ of MILC$_c$ $\beta_{imp}=6.76$ data is shown in Fig. \ref{dalp676} and that of MILC$_f$ is shown in Fig. \ref{dalp709}.


Although statistics is not large, running coupling of CP-PACS and MILC 
show a deviation from pQCD at $q\sim 3$GeV region and the deviation suggests a presence of $A^2$ condensates, gluon condensates and/or quark condensates. The origin of the enhancement of the running coupling of MILC$_f$ in 3-5GeV region is the enhancement of the ghost propagator in $qa>1$ region (Fig.\ref{ghdlmilc}).
Although $\chi^2$ becomes smaller for $c_3\ne 0$, we need further study for verifying presence of condensates of dimension 6. 

\begin{table}[htb]
\beforetab
\begin{tabular}{c c c c c c}
\firsthline
$\beta_{imp}$  & $c_1$ & $c_2$ & $c_3$ & $d$ & $\chi^2/d.o.f$ \\
\midhline
6.76 & 4.18& -2.45 & $0_{fix}$ & -0.0143 & 0.95\\
     & 6.50& -11.7 & 7.47  & -0.0289 & 0.44 \\
\midhline
7.11 &4.27 &  -2.28 & $0_{fix}$ & -0.0281 & 2.6 \\
     &6.58 & -13.05 & 8.47   & -0.0395 & 1.7 \\
\midhline
7.09 &4.29 & -2.30  & $0_{fix}$& -0.0344 & 3.6\\
     &6.58 & -13.84 & 9.27  & -0.0449 & 2.6\\
\lasthline
\end{tabular}
\aftertab
\captionaftertab[]{The $\chi^2$ fit of coefficients $c_{i}$ in $\Delta \alpha_s$ of the MILC$_f$ $\beta_{imp}=7.11$ and 7.09. $c_3$ fixed to 0 and unfixed cases. $d.o.f$ means the degrees of freedom. }\label{cndnsts}
\end{table}

In ref.\cite{davies}, the running coupling of the MILC configuration was measured by using a perturbative expansion for the plaquette and upsilon spectroscopy to set the scale. They found $\alpha_s(8.2$GeV)$\sim 0.214$ at $N_f=2+1$, $m_\pi/m_\rho\sim 0.4$.
 When we fix the scale of MILC$_f$ $\beta_{imp}=7.09$ by the $N_f=3$ pQCD result at $q=$6.84GeV,  $\alpha_s(6.84$GeV)=0.219, our data  $\alpha_s($8.2GeV)=0.190(2) is smaller than \cite{davies} by about 10\%. A possible origin of the difference is that the pQCD results correspond to those of the chiral limit and does not fix the proper scale of the lattice. A more rigorous scale fixing would be achieved by fitting lattice data in high energy region in perturbation series of the running coupling, fixing the  $\alpha_s$ in e.g. V-scheme and converting to the $\overline{\rm MS}$ scheme\cite{davies,davies1,mass}. 

The V-scheme would not be applicable to the infrared region due to the sensitivity to the test-charge wave function. Brodsky et al.\cite{bro} applied physical $\alpha_\tau$ scheme in hypothetical $\tau$ lepton decay and, by using the $\beta_\tau$ three-loop results, observed freezing of the running coupling to the infrared fixed point $\alpha_\tau(0)\sim 2$.  

\section{Conclusion and discussion}
We measured the running coupling and the Kugo-Ojima parameter from unquenched QCD 
gauge configurations of Wilson fermion and KS fermion. 
We observed a sign of different behaviors of the Wilson fermions and the KS fermions as the
system approaches to the continuum limit and the chiral limit. In the case
of the Wilson fermion, the running coupling increases as the mass of the sea quarks decreases and approaches to the chiral limit, while in the case of the KS fermion
dependence on the mass of the sea quark is very weak but it depends on the lattice spacing $a$. 
The $a$ dependence of the KS fermion was expected
to be due to the presence of violation of the taste symmetry which is of order
$a^2$ and the Asqtad action improved this deficiency.  Despite these differences, the running coupling of the Wilson fermion of the smallest quark mass i.e. CP-PACS ($\beta=6.85$) configuration and that of the KS fermion of the  MILC$_f$ and MILC$_c$ ($\beta_{imp}=6.76$) are consistent. Milder finite size effects in Wilson fermions than in KS fermions, due to the spread of the KS fermion over a hypercube in the spinor-flavor interpretation is observed also in \cite{aoki}.

The $c_1/q^2$ correction of the running coupling of $c_1$ of the order of a few GeV is confirmed in the CP-PACS data but not in JLQCD.  Orsay group analyzed the data of Wilson fermion
of $K_{sea}=0.15$, while we analyzed JLQCD Wilson fermion of $K_{sea}=0.1340$ and find the correction is much smaller.
These results indicate that the term $c_1/q^2$ appears as the system approaches the chiral limit.  The MILC$_f$ data suggests that near the chiral limit there are $c_1/q^2$ and $c_2/q^4$ terms which have different signs. The different sign of the subsequent terms causes worriying whether this expansion converges. We need to increase the statistics for obtaining a definite magnitude of the condensates.

On the physical meanings of the $c_1/q^2$ term, there are several discussions.
The Orsay group interprets this term as an indication of the $A^2$ condensates\cite{Orsay, bllmpr,soro}. The operator $A_\mu^2$ is a dimension-2 operator allowed to have a vacuum expectation value, and it appears in the operator product expansion of the running coupling and in the gluon dressing function. Although it is not gauge invariant, the Landau gauge condition $\partial_\mu A_\mu=0$ is compatible with stationarity of $\langle A_\mu^2\rangle$\cite{guza}. In the context of the maximal abelian gauge, the on-shell BRST invariant mixed gluon-ghost condensate of dimension-2 is discussed as gauge invariant observable\cite{Gri,Kondo1}. At tree level the parameter $\tilde c_1$ in the gluon dressing function and $c_1$ in the running coupling in the triple gluon vertex are related to the $\langle A^2\rangle$ as\cite{bllmpr}
\begin{equation}
\tilde c_1=3 g^2\frac{\langle A^2\rangle_{prop}}{4(N_c^2-1)}
\end{equation}
and
\begin{equation}
c_1=9 g^3\frac{\langle A^2\rangle_{alpha}}{4(N_c^2-1)}
\end{equation}
In the ghost anti-ghost gluon vertex, the multiplicity of $c_1$ is reduced
by a factor 3. Thus the fit of running coupling and the gluon dressing function with $\mu=1.97$GeV, $c_3=0$ ansatz yields
$
\displaystyle \frac{\langle A^2\rangle_{alpha}^{1/2}}{\langle A^2\rangle_{prop}^{1/2}}=0.78
$ 
and the corresponding case with $c_3\ne 0$ ansatz yields 0.97.  Orsay group obtained the ratio in the quenched simulation as 1.21\cite{bllmpr} in the 3-loop calculation.

 In the restriction of the gauge field of the Landau gauge in the fundamental modular region, Zwanziger\cite{Zw2} defined the horizon function as $\langle H \rangle=V(N_c^2-1) (d h+e)$, where $V$ is the lattice volume and $d, e$ and $h$ are defined following Eq.(\ref{horizon}). He proposed a simulation with a Boltzmann weight of $e^{-\gamma H}$, where $\gamma^{1/2}$ is a parameter of dimension-2. The dimensional parameter in action breaks the dilatation invariance and it has a link to the global properties of the fundamental modular region. This conjecture was recently discussed in ref.\cite{dssv} including a possible condensation of $A^2$. In this theory Zwanziger's $\gamma$ is affected by the presence of $A^2$ condensates. 
 Since our simulation is done without the Boltzmann weight, we cannot measure the parameter $\gamma$ directly. We observed, however, the horizon function $h$ is negative in quenched simulations and consistent with 0 in unquenched large lattices simulations. The running coupling of the quenched as well as unquenched simulation shows that $\langle A^2\rangle$ is positive. The running coupling of MILC$_f$ suggests that $c_2$ where the gluon condensates $\langle g^2 F_{\mu\nu}^2\rangle$ and/or the quark condensates $\langle m\bar q q\rangle$ contribute is negative. In an analysis of QCD gap equation\cite{dssv}, a solution with negative $\langle g^2 F_{\mu\nu}^2\rangle$, positive $\langle A^2\rangle$ was found. Since the gluon condensates $\langle g^2 F_{\mu\nu}^2\rangle$ and the vacuum energy $E_{vac}$ have opposite signs, it implies that the sign of vacuum energy is positive in contradiction to the result of a two-loop analytical calculation\cite{vkvv}.  

Since $c_2$ term appears only in the running coupling of MILC$_f$ and MILC$_c$ of light sea quark mass, it would be natural to interpret that the $c_2$ term comes mainly from quark condensates. Although the sea-quark mass of MILC$_f$ $m_u+m_d=2\times 0.068 $GeV is too heavy to discuss the chiral limit,  $\langle \bar q q\rangle$ has the correct sign as the Gell-Mann, Oakes, Renner relation\cite{GOR,Nar}
\begin{equation}
{m_\pi}^2f_\pi^2\simeq -(m_u+m_d)\langle \bar q q\rangle+O(m_{ud}^2)
\end{equation}
requires. An analysis of quark propagator\cite{bhlpwz} in Landau gauge also suggests that it is negative.

Momentum dependence of the running coupling in the infrared region is chracterized by the infrared exponent of the gluon dressing function and the ghost dressing function. 
The infrared exponent $\alpha_G$ of the ghost dressing function at 0.4GeV region is about half of $\kappa$ used in the Dyson-Schwinger approach at the infrared limit. We observe $2\alpha_G+\alpha_D\sim 0$ which supports the presence of infrared fixed point\cite{vSHA1}. In the asymmetric lattice we observed that $\alpha_G$ near the lowest momentum along the long lattice axis is suppressed. As an analysis of Dyson-Schwinger equation suggests, this suppression coud be due to the compactness of the lattice\cite{fga}. We suspect, however, there are effects due to the fluctuation of the ghost propagator\cite{FN06}, and there are an elaborated structure of the fixed point which is veiled by Gribov copies.

Zwanziger\cite{Zw1} argued that stochastic gauge fixing would render configurations to the common boundary of the fundamental modular region and the Gribov region and the Gribov copy effects can be evaded.@The argument is based on the assumption that the renormalization group flow of the ghost propagator follows perturbative renormalization-group flow equation. Our simulations of the ghost propagator\cite{FN06} do not confirm the simple renormalization-group flow.  The scenario of suppression of the infrared modes of the gauge field due to vanishing gluon propagator is not confirmed neither and yet to be further investigated.

We admit that we could not restrict our gauge fixed configurations of unquenched simulations in the fundamental modular region. 
In the case of SU(2), we performed parallel tempering(PT) gauge fixing and observed that the ghost propagator of PT gauge fixed samples is less singular than that of the first copy\cite{FN03,NF05}.  In SU(3) of large lattice volume, we found exceptional copies whose $A^2$ norm is larger than the average and the ghost propagator is more singular than the average. These data suggest that infrared features of Gribov copies are complicated.

 Freezing of the running coupling in the infrared is assumed in a model of dynamical chiral symmetry breaking\cite{higashi, Blo1}. A Dyson-Schwinger approach predicts that the infrared fixed point corresponding to $\kappa=0.5$ is about 2.5\cite{Blo1} and the lattice results would not be inconsistent with this model, if the lattice artefacts could be properly removed.

The Kugo-Ojima parameter of MILC configuration is consistent with 1 in the average of polarization, but the value for polarization in t direction is small, since the lattice length transverse to t direction is short in asymmetric lattices. The slope of the ghost propagator $\alpha_G$ also depends on the length of the axis. Differences
between two directions provide a warning on lattice artefacts in the infrared region, however the qualitative difference of the quenched simulation ($c\sim 0.8$) and the unquenched ($c\sim 1$) would be related to the larger fluctuation of the ghost propagator in the quenched simulation\cite{FN06}.

\begin{acknowledge}
We thank the JLQCD collaboration and the CP-PACS collaboration for providing us their unquenched SU(3) lattice configurations. Thanks are also due to MILC collaboration and Columbia university group for the supply of their gauge configurations in the ILDG data base. We are grateful to Taichiro Kugo for valuable discussions and David Dudal for illuminating discussion on horizon function. This work is supported by the KEK supercomputing project 04-106. H.N. is supported by the MEXT grant in aid of scientific research in priority area No.13135210.
\end{acknowledge}

\appendix
\section*{Appendix : The gluon propagator, ghost propagator and the running coupling in pQCD}
\setcounter{section}{1}
In this appendix, we present the pQCD results of the gluon propagator and the ghost propagator and corresponding dressing functions in the $\widetilde{MOM}$ scheme, that are used in fitting the lattice data. We also present the pQCD definition of the QCD running coupling in the $\widetilde{MOM}$ scheme.

The running coupling of the QCD satisfies the renormalization group equation
\begin{equation}
q^2\frac{\partial h}{\partial q^2}=-\beta_0 h^2-\beta_1 h^3-\beta_2 h^4+\cdots
\end{equation}
where in general $h$ is scheme and scale dependent. In the high energy region the $\beta$ function of the $\overline{MS}$ scheme is well behaved and expansion in $h_{\overline{\rm MS}}$ converges,  but in the low energy region, expansion in $h_{\overline{\rm MS}}$ is not a good converging series. We adopt the $\widetilde{MOM}$ scheme 
and express its running coupling in a series of the running coupling of the ${\overline{\rm MS}}$ scheme. The method is known as the effective charge method\cite{grun}.  We define as \cite{vanacol}
 an expansion parameter  $y_{\overline{\rm MS}}(q)$ that satisfies asymptotically 
\begin{equation}
1/y_{\overline{\rm MS}}(q)={\beta_0 \log(q^2/\Lambda_{\overline{\rm MS}}^2)-\frac{\beta_1}{\beta_0}\log(\beta_0 y_{\overline{\rm MS}}(q))}
\end{equation}
in terms of the effective coupling in $\widetilde{MOM}$ scheme $y$ which is a solution of 
\begin{equation}
\frac{1}{y}=\beta_0\log(\mu^2/\Lambda_{\widetilde{MOM}}^2)-\frac{\beta_1}{\beta_0}\log(\beta_0 y)
\end{equation}
Using the function $k(q^2,y)$ defined as
\begin{equation}
k(q^2,y)=\frac{1}{y}+\frac{\beta_1}{\beta_0}\log(\beta_0 y)-\beta_0\log(q^2/\Lambda_{\overline{\rm MS}}^2)
\end{equation}
the expansion parameter $y_{\overline{\rm MS}}(q)$ is expressed as
\begin{eqnarray}
&&y_{\overline{\rm MS}}(q)=y[1+yk(q^2,y)+y^2(\frac{\beta_1}{\beta_0}+k(q^2,y)^2)\nonumber\\
&&+y^3(\frac{{\beta_1}^2}{{\beta_0}^2}k(q^2,y)+\frac{5\beta_1}{2\beta_0}k(q^2,y)^2+k(q^2,y)^3)+\cdots
\end{eqnarray}

In terms of the $y_{\overline{\rm MS}}(q)$, the solution of the renormalization group equation
\begin{eqnarray}
\beta_0\log\frac{q^2}{\Lambda^2}=\frac{1}{h}+\frac{\beta_1}{\beta_0} \log(\beta_0 h)\qquad\qquad\qquad\qquad\qquad\nonumber\\
+\int_0^h dx (\frac{1}{x^2}-\frac{\beta_1}{\beta_0 x}-\frac{\beta_0}{\beta_0 x^2+\beta_1 x^3+\cdots+\beta_n x^{n+2}})\label{heq}
\end{eqnarray}
 can be expressed as
\begin{eqnarray}\label{hmsbar}
h(q)&=&y_{\overline{\rm MS}}(q)(1+y_{\overline{\rm MS}}(q)^2(\bar\beta_2/\beta_0-(\beta_1/\beta_0)^2)\nonumber\\
&&+y_{\overline{\rm MS}}(q)^3\frac{1}{2}(\bar \beta_3/\beta_0-(\beta_1/\beta_0)^3)+\cdots
\end{eqnarray}
where
$\displaystyle \beta_0=11-\frac{2}{3}N_f, \beta_1=102-\frac{38}{3}N_f$,
are scheme independent, and in the $\overline {\rm MS}$ scheme

\noindent$\displaystyle \bar\beta_2 = \frac{2857}{2} - \frac{5033}{18} N_f + \frac{325}{54} N_f^2$,

\noindent$\displaystyle \bar\beta_3 = (\frac{149753}{6} + 
          3564 \zeta(3) + (-\frac{1078361}{162} N_f - 
              \frac{6508}{27} N_f \zeta(3))$.

In the effective charge method, the propagator $D^{ab}_{\mu\nu}(-q^2)$ (we use Minkovski metric here) is expressed by the scale and scheme invariant propagator $\hat D^{ab}_{\mu\nu}(-q^2)$ and a function $f(h)$\cite{vanacol,chet}
\begin{equation}
\hat D^{ab}_{\mu\nu}(-q^2)=f(h)D^{ab}_{\mu\nu}(-q^2).
\end{equation}
Where $D^{ab}_{\mu\nu}$ satisfies the renormalization group equation
\begin{equation}
\mu^2\frac{\partial}{\partial \mu^2}D^{ab}_{\mu\nu}(-q^2)\equiv (\gamma_{0}h+\gamma_{1}h^2+\gamma_{2}h^3+\cdots)D^{ab}_{\mu\nu}(-q^2)
\end{equation}
and
\begin{equation}\label{fh}
f(h)=exp\int^h \frac{dx}{x}\frac{\gamma(x)}{\beta(x)}
\end{equation}
The general solution of (\ref{fh}) is
\begin{eqnarray}
&&f(h)=\lambda h^{\bar\gamma_{0}}[1+(\bar\gamma_{1}-\bar\gamma_{0}\bar\beta_1')h\nonumber\\
&&+\frac{1}{2}((\bar\gamma_{1}-\bar\beta_1'\bar\gamma_{0})^2+\bar\beta_2'\bar\gamma_{0}+\bar\beta_1'^2\bar\gamma_0\nonumber\\
&&-\bar\beta_1'\bar\gamma_{1}-\bar\beta_2'\bar\gamma_{0})h^2\nonumber\\
&&+(\frac{1}{6}(\bar\gamma_{1}-\bar\beta_1'\bar\gamma_{0})^3
+\frac{1}{2}(\bar\gamma_{1}-\bar\beta_1'\bar\gamma_{0})\nonumber\\
&&(\bar\gamma_{2}+\bar\beta_1'^2\bar\gamma_{0}-\bar\beta_1'\bar\gamma_{1}-\bar\beta_2'\bar\gamma_{0})\nonumber\\
&&+\frac{1}{3}(\bar\gamma_{3}-\bar\beta_1'^3\bar\gamma_{0}+2\bar\beta_1'\bar\beta_2'\bar\gamma_{0}-\bar\beta_3'\bar\gamma_{0}+\bar\beta_1'^2\bar\gamma_{1}
\nonumber\\
&&-\bar\beta_2'\bar\gamma_{1}-\bar\beta_1'\bar\gamma_{2})h^3+\cdots]
\end{eqnarray}
where
$\displaystyle\bar\beta_i'=\frac{\bar\beta_i}{\beta_0}$ ($i=1,2,3$) and 
 $\displaystyle\bar\gamma_{j}=\frac{\gamma_{j}}{\beta_0}$ ($j=0,1,2,3$).

\subsection{Gluon dressing function}

The anomalous dimension $\gamma_{3}$  of the gluon propagator is
\[
\gamma_3=\gamma_{3_0}h+\gamma_{3_1}h^2+\gamma_{3_2}h^3+\gamma_{3_3}h^4+\cdots\]
where

 $\displaystyle\gamma_{3_0}=\frac{13}{2}-\frac{2N_f}{3}$,

 $\displaystyle \gamma_{3_1}=9\frac{59}{8}-N_f\frac{15}{2}-N_f\frac{32}{12}$,

\begin{eqnarray}
&&\gamma_{3_2}=27(\frac{9965}{288}-\frac{9}{16} \zeta(3))\nonumber\\
&&+\frac{9}{2}N_f(-\frac{911}{18}+18 \zeta(3))+2N_f(-\frac{5}{18}-24 \zeta(3))\nonumber\\
&&+\frac{76}{12} N_f^2+\frac{N_f^2}{3}\frac{44}{9}+\frac{16}{9}N_f
\end{eqnarray}
 and
\begin{eqnarray}
&&\gamma_{3_3}=-(-\frac{10596127}{768}+\frac{1012023}{256} \zeta(3) -\frac{8019}{3} \zeta(4)\nonumber\\
&& -\frac{40905}{4} \zeta(5)\nonumber\\
         &&+N_f(\frac{23350603}{5184}-\frac{387649}{216}\zeta(3)+\frac{8955}{16}\zeta(4)+\frac{3355}{2}\zeta(5))\nonumber\\
          &&+N_f^2(-\frac{43033}{162}-\frac{2017}{81}\zeta(3) -33\zeta(4))\nonumber\\
&&+N_f^3(-\frac{4427}{1458}+\frac{8}{3} \zeta(3)))
\end{eqnarray}

Using the above gluon field anomalous dimension of four-loop level in $\widetilde{MOM}$ scheme \cite{chet} and the coupling constant $h(q)$ in $\overline{\rm MS}$ scheme (\ref{hmsbar}), which is a function of the parameter $y$ defined in the $\widetilde{MOM}$ scheme,
we derive the gluon propagator $D_A(-q^2)$ in the $\widetilde{MOM}$ scheme as a solution of the renormalization group equation
\begin{equation}
\mu^2\frac{\partial}{\partial \mu^2}D_A(-q^2)\equiv (\gamma_{3_0}h+\gamma_{3_1}h^2+\gamma_{3_2}h^3+\cdots)D_A(-q^2)
\end{equation}

The gluon dressing function $Z(q^2)=q^2 D_A(q^2)$ in Eucledian metric becomes
\begin{eqnarray*}
&&Z^{-1}=\lambda^{-1}h^{-\frac{39-4n}{66-4n}}[1-\frac{3 \left(104 n^2-1974 n+15813\right) h}{16
   (33-2 n)^2}\nonumber\\
&&+\{
(\left(128000 n^5-192 (53419+504
   \zeta (3)) n^4\right.\nonumber\\
&&+288 (1235761+5238 \zeta (3)) n^3\nonumber\\
&&+108
   (-56578007+772200 \zeta (3)) n^2\nonumber\\
&&-324
   (-153696523+6930396 \zeta (3)) n\nonumber\\
&&+\left. 243
   (-615512003+60661656 \zeta (3))\right) h^2\} \nonumber\\
&&/ (1536
   (33-2 n)^4) \nonumber\\
&&+\{\left(-354549760 n^8-663552
   (-79985+304 \zeta (3)) n^7\right.\nonumber\\
&&+4608
   (-738019369+10620024 \zeta (3)+4370400 \zeta (5))
   n^6\nonumber\\
&&-3456 (-34931893063+1008068136 \zeta
   (3)\nonumber\\
&&+523278720 \zeta (5)) n^5\nonumber\\
&&+2592 (46615708836
   \zeta (3)+275 (-3654988631\nonumber\\
&&+93932928 \zeta (5)))
   n^4\nonumber\\
&&-5832 (-6037451357147+404411943104 \zeta
   (3)\nonumber\\
&&+223742745600 \zeta (5)) n^3\nonumber\\
&&+2916
   (-100416325711969+9138430613136 \zeta
   (3)\nonumber\\
&&+4824029548800 \zeta (5)) n^2\nonumber\\
&&-4374
   (-314978703784231+37405611077472 \zeta
   (3)\nonumber\\
&&+18085875033600 \zeta (5)) n\nonumber\\
&&+ 6561
   (-430343889400537+64653527897640 \zeta
   (3)\nonumber\\
&&+\left.27401036762880 \zeta (5))\right) h^3\}\nonumber\\
&&/(663552
   (33-2 n)^6) \nonumber\\
&&+O\left(h^4\right)]
\end{eqnarray*}
where $n=N_f$.

\subsection{The ghost dressing function}

The anomalous dimension $\tilde\gamma_{3}$  of the ghost propagator is
\[
\tilde\gamma_3=\tilde\gamma_{3_0}h+\tilde\gamma_{3_1}h^2+\tilde\gamma_{3_2}h^3+\tilde\gamma_{3_3}h^4+\cdots\]
where

$\displaystyle\tilde\gamma_{3_0}=\frac{9}{4}$,

$\displaystyle \tilde\gamma_{3_1}=9\frac{95}{48}-3N_f\frac{5}{12}$, 
\begin{eqnarray} 
&&\tilde\gamma_{3_2}=2N_f(-\frac{45}{4}+12\zeta(3))+\frac{3}{4}N_f^2(-\frac{35}{27})\nonumber\\
&&+\frac{9}{2}N_f(-\frac{97}{108}-9\zeta(3))+27(\frac{15817}{1728}+\frac{9}{32}\zeta(3))\nonumber
\end{eqnarray}
and
\begin{eqnarray}
 &&\tilde\gamma_{3_3}=-(-\frac{2857419}{512}-\frac{1924407}{512} \zeta(3)+\frac{8019}{64} \zeta(4)\nonumber\\
&&+\frac{40905}{8} \zeta(5)+
          N_f(\frac{1239661}{1152}+\frac{48857}{48} \zeta(3)-\frac{8955}{32} \zeta(4)\nonumber\\
&&-\frac{3355}{4} \zeta(5))\nonumber\\
          &&+N_f^2(-\frac{586}{27}-\frac{55}{2}\zeta(3)+\frac{33}{2} \zeta(4))\nonumber\\
&&+N_f^3(\frac{83}{108}-\frac{4}{3} \zeta(3)))
\end{eqnarray}

Using the ghost field anomalous dimension of four-loop level in $\widetilde{MOM}$ scheme and the coupling constant $h(q)$ in $\widetilde{MOM}$ scheme\cite{chet}, we derive the ghost propagator $D_G(-q^2)$ in the $\widetilde{MOM}$ scheme as a solution of the renormalization group equation
\begin{equation}
\mu^2\frac{\partial}{\partial \mu^2}D_G(-q^2)\equiv (\tilde\gamma_{3_0}h+\tilde\gamma_{3_1}h^2+\tilde\gamma_{3_2}h^3+\cdots)D_G(-q^2)
\end{equation}

The ghost dressing function $G(q^2)=q^2 D_G(q^2)$ in Eucledian metric becomes
\begin{eqnarray}
&&G^{-1}={\lambda_G}^{-1} h^{-\frac{27}{132-8n}}[1+\nonumber\\
&&\left(\frac{10 n}{9}+\frac{3 \left(40 n^2+138
   n-1611\right)}{8 (33-2 n)^2}-\frac{97}{12}\right)h\nonumber\\
&&+\{\left(512 (1439+48 \zeta (3)) n^5\right.\nonumber\\
&&-3840(13883+174 \zeta (3)) n^4\nonumber\\
&&-864 (-1728454+17931 \zeta(3)) n^3\nonumber\\
&&+108 (-185691691+6984516 \zeta(3)) n^2\nonumber\\
&&-324(-395301865+28831638 \zeta (3)) n\nonumber\\
&&+\left. 19683(-15277259+1921964 \zeta (3))\right) h^2\}\nonumber\\
&&/ (1152(33-2 n)^4)\nonumber\\
&&+\{\left(-16384 (174163+432 \zeta(3)) n^8\right.\nonumber\\
&&-12288 (-30802025+637212 \zeta (3)+1002240 \zeta (5)) n^7\nonumber\\
&&+4608 (-4614333119+207142932 \zeta(3)\nonumber\\
&&+266137650 \zeta (5)) n^6\nonumber\\
&&-3456
   (-192809757953+13588881045 \zeta (3)\nonumber\\
&&+14877266760
   \zeta (5)) n^5\nonumber\\
&&+5184 (-2470563836117+240877496568
   \zeta (3)\nonumber\\
&&+225684805500 \zeta (5)) n^4\nonumber\\
&&-1944
   (-79693953595001+10028539488648 \zeta(3)\nonumber\\
&&+7956577252800 \zeta (5)) n^3\nonumber\\
&&+43740
   (-26302376345491+4087102826048 \zeta(3)\nonumber\\
&&+2675346352272 \zeta (5)) n^2\nonumber\\
&&- 13122
   (-363568314295693+67715969212716 \zeta(3)\nonumber\\
&&+34697940156000 \zeta (5)) n\nonumber\\
&&+59049
   (-141629801206331+31037533417440 \zeta(3)\nonumber\\
&&+\left.11069576361360 \zeta (5))\right) h^3\} \nonumber\\
&&/(746496 (33-2 n)^6) \nonumber\\
&&+O\left(h^4\right)].
\end{eqnarray}

\subsection{The running coupling}
In perturbative QCD, running coupling is derived by the renormalization group equation 
\begin{equation}
\frac{\partial \alpha}{\partial\log\mu}=-(\frac{\beta_0}{2\pi}\alpha^2+\frac{\beta_1}{4\pi^2}\alpha^3+\frac{\beta_2}{64\pi^3}\alpha^4+\frac{\beta_3}{128\pi^4}\alpha^5)+o(\alpha^6)
\end{equation}

In the $\widetilde{MOM}$ scheme, inversion of the 2-loop formula
\begin{equation}
\Lambda=\mu e^{-\frac{2\pi}{\beta_0 \alpha_s}}(\frac{\beta_0\alpha_s}{4\pi})^{-\beta_1/\beta_0^2}
\end{equation}
can be done analytically with use of the Lambert W function that satisfies $z=W[z]e^{W[z]}$. We find
\begin{equation}
\frac{\alpha_s}{2\pi}=\frac{\beta_0}{\beta_1} W[(\beta_0^2/2\beta_1)e^{(\beta_0^2 /2\beta_1)t}]
\end{equation}
where $t=\log(\mu^2/\Lambda^2)$.

An approximate inversion of the four-loop formula yields the running coupling as a function of $t=\log(\mu^2/\Lambda^2)$ as follows\cite{Orsay3,ChRe}.

\begin{eqnarray}
&&\alpha_{s,pert}(\mu)=\frac{4\pi}{\beta_0 t}-\frac{8\pi\beta_1}{\beta_0}\frac{\log (t)}{(\beta_0 t)^2}\nonumber\\
&&+\frac{1}{(\beta_0t)^3}\left(\frac{2\pi \beta_2}{\beta_0}+\frac{16\pi\beta_1^2}{\beta_0^2}(\log^2(t)-\log(t)-1)\right)\nonumber\\
&&+\frac{1}{(\beta_0t)^4}\left[\frac{2\pi\beta_3}{\beta_0}+\frac{16\pi\beta_1^3}{\beta_0^3}\left(-2\log^3(t)+5\log^2(t)\right.\right.\nonumber\\
&&\left.\left.+(4-\frac{3\beta_2\beta_0}{4\beta_1^2})\log(t)-1\right)\right]
\end{eqnarray}


\begin{thebibliography}{99}
\bibitem{Zw2} Zwanziger, D. :{\NPB}{\bf 399},{477}{(1993)}. The horizon function $\langle H\rangle$ of this paper is different from that of \cite{Zw} and ours by the addition of $eV(N_c^2-1)$.
\end{thebibliography}
\end{document}